\documentclass[prd,superscriptaddress, nofootinbib, preprintnumbers]{revtex4-2}
\usepackage{amsmath}
\usepackage{mathtools}
\usepackage{graphicx}
\usepackage{xcolor}
\usepackage{hyperref}

\usepackage{float}
\usepackage{braket}
\usepackage{comment}

\newcommand{\be}{\begin{equation}}
	\newcommand{\ee}{\end{equation}}

\begin{document}
	\title{An accurate fluid approximation for massive neutrinos in cosmology}
	\author{Caio Nascimento \\{\it{\small  Department of Physics, University of Washington, Seattle, WA, USA}}}

\begin{abstract}

A measurement of the neutrino mass scale will be achieved with cosmological probes in the upcoming decade. On one hand, the inclusion of massive neutrinos in the linear perturbation theory of cosmological structure formation is well understood and can be done accurately with state of the art Boltzmann solvers. On the other hand, the numerical implementation of the Boltzmann equation is computationally expensive and is a bottleneck in those codes. This has motivated the development of more efficient fluid approximations, despite their limited accuracy over all scales of interest, $k \sim (10^{-3}-10)$Mpc$^{-1}$. In this work we account for the dispersive nature of the neutrino fluid, i.e., the scale dependence in the sound speed, leading to an improved fluid approximation. We show that overall $\lesssim 5\%$ errors can be achieved for the neutrino density and velocity transfer functions at redshift $z \lesssim 5$, which corresponds to an order of magnitude improvement over previous approximation schemes that can be discrepant by as much as a factor of two.

\end{abstract}
	
\maketitle

\bibliographystyle{unsrt}
	
\section{Introduction}

The observation of neutrino oscillations has established that at least two of the neutrino mass eigenstates have a non-zero mass, with an associated lower bound on the sum of the masses of $\sum_{\nu} m_{\nu} \gtrsim 0.06$ and $0.1$eV for the normal and inverted hierarchies respectively \cite{de20212020,capozzi2020addendum,esteban2020fate}. Complementary information comes from beta decay experiments, which set an upper bound to a weighted sum of the masses $m_{\nu, \beta} < 0.8$eV \cite{aker2022katrin}. Additionally, massive neutrinos suppress cosmological structure formation at small scales \cite{lesgourgues2006massive}, leading to the most stringent upper bound on the sum of neutrino masses to date, i.e., $\sum_{\nu} m_{\nu} < 0.12$eV \cite{collaboration2020planck}. \footnote{This upper bound can be relaxed with nonstandard scenarios such as an unstable neutrino species and dynamical dark energy \cite{hannestad2005neutrino,escudero2020relaxing,chacko2020cosmological}.} It is expected that future cosmological surveys in the upcoming decade will be sensitive to the lower bound from oscillation experiments and hence will allow for a detection of the neutrino mass scale \cite{aghamousa2016desi,abell2009lsst,abazajian2016cmb}. This is a crucial measurement since it sets a clear target for laboratory experiments and serves as a cross-check of the consistency between particle physics and cosmology \cite{green2021cosmological}.

The inclusion of massive neutrinos in linear cosmological perturbation theory has a long history (see \cite{ma1995cosmological} and references therein). Due to the large velocity dispersion of massive neutrinos, one must go beyond a simple fluid treatment and solve a hierarchy of Boltzmann equations in phase space. At each time step the neutrino distribution function is then integrated over momenta to obtain the neutrino stress-energy tensor, which in turn contributes to the right hand side of Einstein's equations and sets the coupling of neutrinos to the other species in the universe. This is a cumbersome procedure and a computational bottleneck in state of the art Boltzmann solvers, such as the Cosmic Linear Anisotropy Solving System (CLASS) \cite{lesgourgues2011cosmic} and the Code for Anisotropies in the Microwave Background (CAMB) \cite{lewis2017code}. 

As a consequence, the search for more efficient alternative approaches to the inclusion of massive neutrinos in linear cosmological perturbation theory remains a well motivated direction of research since modern cosmological parameter inference techniques require these codes to be run tens or hundreds of thousands of times. For instance, \cite{ji2022cosmological} formulates the problem as an integral equation and proposes an iterative solution, while \cite{de2021generalized} integrates out the momentum dependence at the cost of a significant increase in the dimensionality of the resulting system of ordinary differential equations. 

An alternative approach consists of a simple fluid approximation for the exact neutrino dynamics. This is a viable option whenever the neutrinos only give a small contribution to the total matter energy density, since we can then afford for some inaccuracies in the neutrino density provided we are only interested in the total matter (or cold dark matter) field. This is especially true on small scales where the neutrino density is suppressed due to free-streaming and the cold dark matter evolution basically decouples from the neutrinos, the same circumstances in which the Boltzmann hierarchy needs to be truncated at a large multipole and becomes computationally expensive. Indeed, a fluid approximation is used in CLASS to evolve the neutrino component on scales that are much smaller than the cosmological horizon \cite{lesgourgues2011cosmic}. Another fluid approximation for massive neutrinos follows from the generalized dark matter approach of \cite{hu1998structure}. 

What all fluid approximations have in common is that they become inaccurate at sufficiently small scales \cite{shoji2010massive}, exactly in the regime where the approximation is the most useful since the exact dynamics is more (computationally) expansive as we discussed above. In this work we show that this failure of the fluid approximation at small scales is mostly a result of \textit{not} accounting for the dispersive nature of the neutrino fluid, i.e., the sound speed is scale dependent \cite{inman2016cosmic}. In previous works much of the focus was directed towards modeling the evolution of the neutrino shear stress implicitly presuming that the assumption of an adiabatic sound speed does not dominate the total error \cite{lesgourgues2011cosmic}. Instead, we find that the assumption of an adiabatic sound speed leads to a significant overestimation of the sound speed on small scales that dominates the error in the fluid approximation.

We obtain a simple analytic expression for the sound speed at small scales and use it to introduce a scale dependent approximation to this quantity that interpolates between the small and large scale regimes. This, in combination with a scale dependent approximate expression to the anisotropic stress, leads to a resulting fluid approximation for massive neutrinos with $\lesssim 5\%$ errors for the neutrino density and velocity transfer functions at redshift $z\lesssim 5$ and over scales $k=(10^{-3}-10)$Mpc$^{-1}$, which corresponds to an order of magnitude improvement over previous approximation schemes that can be as much as a factor of two wrong.

We consider neutrino masses in the range $0.02$eV $\leq m_{\nu} \leq 0.5$eV, for which a $\lesssim 5\%$ error in the neutrino component is sufficient to produce the total matter power spectrum to sub-percent level accuracy. This fluid approximation is then a powerful alternative to the full Boltzmann hierarchy for most projects, allowing for a significant reduction in computing time. 

The paper is organized as follows: In Section \ref{sec:fluideqs} we introduce the fluid equations and the approximate expressions for the sound speed and anisotropic stress in the Newtonian gauge. In Section \ref{sec:num} we compare our fluid approximation with the results from CLASS in both high and default precision settings, along with the fluid approximation used in CLASS. In Section \ref{sec:conc} we summarize our results. Details of calculations that motivate the approximations employed can be found in Appendix \ref{sec:app}, and in Appendix \ref{sec:app2} we extend our fluid approximation to alternative gauges (other than the Newtonian gauge), showing explicit expressions in the synchronous gauge.
   
\section{Fluid equations}
\label{sec:fluideqs}

The fluid equations satisfied by massive neutrinos are quite generic as they follow from energy-momentum conservation laws. In this section we first introduce the relevant equations (referring the reader to \cite{ma1995cosmological} for further details). Then we briefly motivate and write down formulas that approximate the scale dependent sound speed and anisotropic stress. Derivations and details can be found in Appendix \ref{sec:app}.

We consider scalar perturbations to the Friedmann-Lema\^itre-Robertson-Walker (FLRW) universe in the (conformal) Newtonian gauge, where the metric reads
\begin{equation}
\label{eq:newgauge}
	ds^2 = a^2(\tau)\left[-(1+2\psi)d\tau^2 + (1-2\phi)d\vec{x}^2\right] \,,
\end{equation}
and where $a(\tau)$ is the scale factor, $\tau$ is the conformal time (related to the cosmic time $t$ via the expression $dt = ad\tau$), $\vec{x}$ are comoving spatial coordinates, $\phi(\tau,\vec{x})$ and $\psi(\tau,\vec{x})$ are  gravitational potentials that we treat as (small) linear perturbations. We also define the conformal Hubble rate $\mathcal{H} = a'/a$, where throughout a prime denotes derivative with respect to conformal time $\tau$.

At the level of background the neutrinos are distributed in phase-space with the relativistic Fermi-Dirac profile, $f_{0}(q) = f_{\textrm{FD}}(q/T_{\nu,0})$, where
\begin{equation}
\label{eq:fermidirac}
	f_{FD}(x) = \frac{g_{\nu}}{e^{x}+1} \,,
\end{equation} 
with $q$ the magnitude of the comoving momentum $\vec{q}$, $T_{\nu,0} \approx 1.95$K $\approx 1.7 \times 10^{-4}$eV is the neutrino temperature today, $x=q/T_{\nu,0}$ and $g_{\nu}=2$ to account for both left-handed neutrinos and right-handed antineutrinos. From the condition of isotropy the only non-vanishing components of the neutrino stress-energy tensor are its energy density and pressure. They can be obtained from Eq.(\ref{eq:fermidirac}) as follows:
\begin{align}
	& \rho(a) = a^{-4} \int_{0}^{\infty} \frac{dq}{2\pi^2} q^2 \epsilon(q,a) f_{0}(q)  \,, \label{eq:fluidprop1}  \\ & P(a) = \frac{1}{3} a^{-4} \int_{0}^{\infty} \frac{dq}{2\pi^2} q^2 \epsilon(q,a)  \left[\frac{q}{\epsilon(q,a)}\right]^2 f_{0}(q) \,. \label{eq:fluidprop2}
\end{align}
Here $\epsilon(q,a) = \sqrt{q^2 +m_{\nu}^2a^2}$ is the comoving energy. Note that these are not the total energy density and pressure of the Universe, as they refer only to the neutrino species.  

It will also prove useful to define the equation of state as:
\begin{equation}
\label{eq:eos}
	w(a) = \frac{P(a)}{\rho(a)} \,.
\end{equation}

In the presence of non-vanishing gravitational potentials in Eq.(\ref{eq:newgauge}), the neutrino density and pressure also acquire perturbations, $\delta \rho (\tau, \vec{x})$ and $\delta P(\tau, \vec{x})$, that are time and position dependent. Additionally, there is a net bulk flow that we parameterize by the divergence in the velocity field $\theta(\tau,\vec{x})$, and an anisotropic (shear) stress $\sigma(\tau,\vec{x})$. We follow the standard practice of working in terms of a density contrast $\delta(\tau,\vec{x}) = \delta \rho (\tau, \vec{x})/\rho(a)$, and define the sound speed as
\begin{equation}
\label{eq:effsp}
	c_{\textrm{s}}^2(\tau,\vec{x}) = \frac{\delta P(\tau, \vec{x})}{\delta \rho (\tau, \vec{x})} \,.
\end{equation}

We now have all the ingredients to write down the fluid equations which are exact and follow from the conservation of the neutrino stress-energy tensor \footnote{Moving forward we work in Fourier space $\vec{\nabla} \to i\vec{k}$ and often omit time and scale dependences of fluid properties for simplicity of notation.},
\begin{align}
& \delta' = -(1+w)(\theta-3\phi') -3\mathcal{H}(c_{\textrm{s}}^2-w)\delta \,, \label{eq:fluid1}\\ & \theta' = - \mathcal{H}(1-3w) \theta -\frac{w'}{1+w} \theta + \frac{c_{\textrm{s}}^2}{1+w}k^2 \delta -k^2 \sigma +k^2 \psi \,. \label{eq:fluid2}
\end{align}

In order to close the system of equations we need approximate expressions for both the sound speed $c_{\textrm{s}}^2$ and the anisotropic stress $k^2\sigma$ (in the full Boltzmann hierarchy they can be obtained from the distribution function after an integration over momentum). A complete discussion on the motivations for our approximations (with relevant derivations) can be found in Appendix \ref{sec:app}. Here we will just introduce the main ideas. At sufficiently large scales $c_{\textrm{s}}^2$ approaches the so-called adiabatic sound speed,
\begin{equation}
	\label{eq:adia}
	c_{\textrm{g}}^2(a) = \frac{P'(a)}{ \rho'(a)} \,.
\end{equation}
This follows from separate universe arguments: At sufficiently large scales the neutrino anisotropies can be absorbed into a local shift of the neutrino temperature, $T_{\nu} \to T_{\nu}(1+\mathcal{N_{\nu}})$, with $\mathcal{N_{\nu}} = \delta T_{\nu}/T_{\nu}$ a constant. It then follows from Eq.(\ref{eq:fermidirac}) that the total distribution function reads,
\begin{equation}
\label{eq:dfls}
	f(q) = f_{\textrm{FD}}\left(\frac{q}{T_{\nu,0}(1+\mathcal{N}_{\nu})}\right) = f_{0}(q)\left(1-\frac{d\ln f_{0}}{d\ln q} \mathcal{N}_{\nu} \right) \implies \delta f = -q \frac{d f_{0}}{dq} \mathcal{N}_{\nu} \,, 
\end{equation}
where from the second to the third line we expanded to leading order in $\mathcal{N_{\nu}} = \delta T_{\nu}/T_{\nu}$. Equation (\ref{eq:adia}) can now be obtained upon integration over the comoving momentum to produce the neutrino density and pressure perturbations. Similarly, at sufficiently small scales $c_{\textrm{s}}^2$ approaches what we call the asymptotic (asp) sound speed,
\begin{equation}
	\label{eq:asp}
	c_{\textrm{asp}}^2(a) = \frac{1}{3} \frac{1+w(a)}{1+\lambda(a)} \,,
\end{equation}
where the quantity $\lambda(a)$ is defined by,
\begin{equation}
	\label{eq:lambda}
	\rho(a)\lambda(a) = \frac{1}{3} a^{-4} \int_{0}^{\infty} \frac{dq}{2\pi^2} q^2 \epsilon(q,a) \left[\frac{\epsilon(q,a)}{q}\right]^2 f_{0}(q) \,.
\end{equation}
Equation.(\ref{eq:asp}) can be extracted from the static limit of the Boltzmann equation leveraging on the following observation: Neutrinos have a velocity $v \equiv q/\epsilon$ so that neutrino fluctuations with a (comoving) wavenumber $k$ have a characteristic time scale, $t \sim a/k v$, which is much smaller than a Hubble time at scales $k \gg \mathcal{H}/v$. We can then consider the static limit where the expansion of the universe can be taken as slow and the general solution to the Boltzmann equation is an arbitrary function of the (adiabatic invariant) total comoving energy. At the  background level this is $\epsilon_{\textrm{tot}}=\epsilon= \sqrt{q^2+m_{\nu}^2a^2}$, and the distribution function is $f(q) = f_{0}(\sqrt{\epsilon^2-m_{\nu}^2a^2})$ as given by Eq.(\ref{eq:fermidirac}). However, in the presence of a gravitational potential $\psi$ the total comoving energy reads $\epsilon_{\textrm{tot}}=\epsilon(1+\psi)$. The total distribution function then becomes,
\begin{equation}
	\label{eq:distribution}
	f=f_{0}(\sqrt{\epsilon^2(1+\psi)^2 -m_{\nu}^2a^2}) = f_{0}(q)\left[1+\frac{d\ln f_{0}}{d\ln q} \left(\frac{\epsilon}{q}\right)^2 \psi\right] \implies \delta f = q\frac{df_{0}}{dq} \left(\frac{\epsilon}{q}\right)^2 \psi
\end{equation}
where we expand to leading order in $\psi$. Equation \ref{eq:asp} can then be derived by integrating over the comoving momenta to obtain the neutrino fluid properties. More details can be found in Appendix \ref{sec:app}, where we systematically derive the Eqs. (\ref{eq:adia}) and (\ref{eq:distribution}) as the large and small scale limits of the Boltzmann equation. 

\begin{figure}
	\centering
	\includegraphics[width=0.75\textwidth]{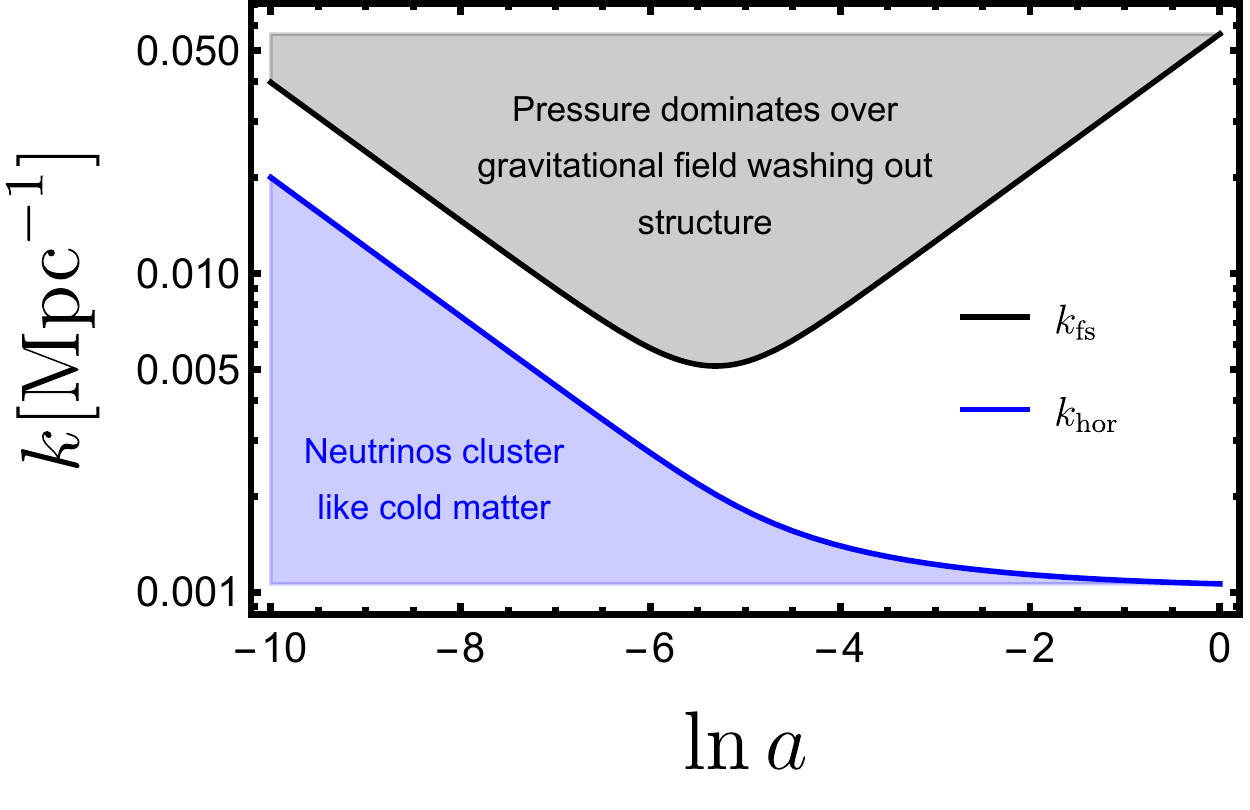}
	\caption{Neutrino free-streaming (black curve) and neutrino horizon (blue curve) scales as a function of the scale factor, for an individual neutrino mass of $m_{\nu}=0.1$eV. The black dashed region corresponds to sub free-streaming scales where pressure dominates over the gravitational field washing out structure, and the blue shaded region corresponds to scales above the neutrino horizon where neutrinos cluster like cold matter. }
	\label{fig:scales}
\end{figure}

Our strategy will now be to interpolate between these two regimes in order to write an approximate expression for the sound speed that accounts for its scale dependence. We similarly also want to introduce a scale dependent approximate expression to the anisotropic stress. To accomplish these goals we first need to understand what are the characteristic scales associated to the neutrino thermal motion.  Indeed, there are two time dependent scales. One is the (instantaneous) free-streaming scale $k_{\textrm{fs}}(a)$ defined by,
\begin{equation}
	\label{eq:fs}
	k_{\textrm{fs}}(a) = \sqrt{\frac{3}{2} \Omega_{\textrm{m}}(a)} \frac{\mathcal{H}(a)}{c_{\textrm{asp}}(a)} \,, 
\end{equation}
which is proportional to the (comoving) distance that neutrinos travel over the course of one expansion time $\tau \sim 1/\mathcal{H}$, i.e., $\lambda_{\textrm{fs}} = 2\pi/k_{\textrm{fs}} \sim c_{\textrm{asp}}/\mathcal{H} \sim  c_{\textrm{asp}}t$, with $\Omega_{\textrm{m}}(a)$ the fractional contribution of matter (including neutrinos) to the total energy budget. Notice from Eqs.(\ref{eq:fluidprop1}), (\ref{eq:fluidprop2}), (\ref{eq:eos}), (\ref{eq:asp}) and (\ref{eq:lambda}) that in the nonrelativistic regime, where $\epsilon \approx ma \gg q$, we have that $\lambda \gg 1$ and $w\ll 1$ such that, 
\begin{equation}
\label{eq:cspnr}
	c_{\textrm{asp}}(a) \approx \frac{1}{\sqrt{3\lambda(a)}} \approx \sigma_{\nu}(a)
\end{equation}
where we have used Eq.(\ref{eq:fermidirac}), and introduce the neutrino velocity dispersion, \footnote{In Appendix \ref{sec:app} we motivate our choice of Eq.(\ref{eq:vds}) for the neutrino velocity dispersion, and hence the appearance of $c_{\textrm{asp}}(a)$ in the definition of the free-streaming scale as in Eq.(\ref{eq:fs}).}
\begin{equation}
	\label{eq:vds}
	\sigma_{\nu}(a) = \sqrt{\frac{3\zeta(3)}{\ln 4}} \frac{T_{\nu,0}}{m_{\nu}a} \,.
\end{equation}
The other scale associated to the neutrino thermal motion is the neutrino horizon, $\lambda_{\textrm{hor}}(a)$, defined by:
\begin{equation}
	\label{eq:hor}
	\lambda_{\textrm{hor}}(a) = \int_{0}^{a} d\ln a' \lambda_{\textrm{fs}}(a') \,,
\end{equation}
or its wavenumber $k_{\textrm{hor}}(a)=2\pi/\lambda_{\textrm{hor}}(a)$, where similarly $\lambda_{\textrm{fs}}(a)=2\pi/k_{\textrm{fs}}(a)$. The neutrino horizon is proportional to the total distance traveled by neutrinos over the entire expansion history. In fact, at late times $k_{\textrm{fs}} \gg k_{\textrm{hor}}$ and both scales play a role in the dynamics of massive neutrinos: $k_{\textrm{hor}}$ is the scale below which ($k \leq k_{\textrm{hor}}$) neutrino velocities can be ignored, and hence neutrinos cluster like cold dark matter, while $k_{\textrm{fs}}$ is the scale above which ($k \geq k_{\textrm{fs}}$) the neutrino pressure dominates over the gravitational potential leading to the suppression of neutrino structure. This is illustrated in Fig.\ref{fig:scales} for an individual neutrino mass of $m_{\nu}=0.1$eV. While the neutrino horizon grows with the expansion of the universe, the free-streaming scale peaks when the neutrinos first become nonrelativistic due to the subsequent decrease in the thermal velocity. This produces a large separation of scales at late times, which is the reason why galaxy surveys cannot directly probe the scale dependence of the neutrino suppression.   

\begin{figure}
	\centering
	\includegraphics[width=0.75\textwidth]{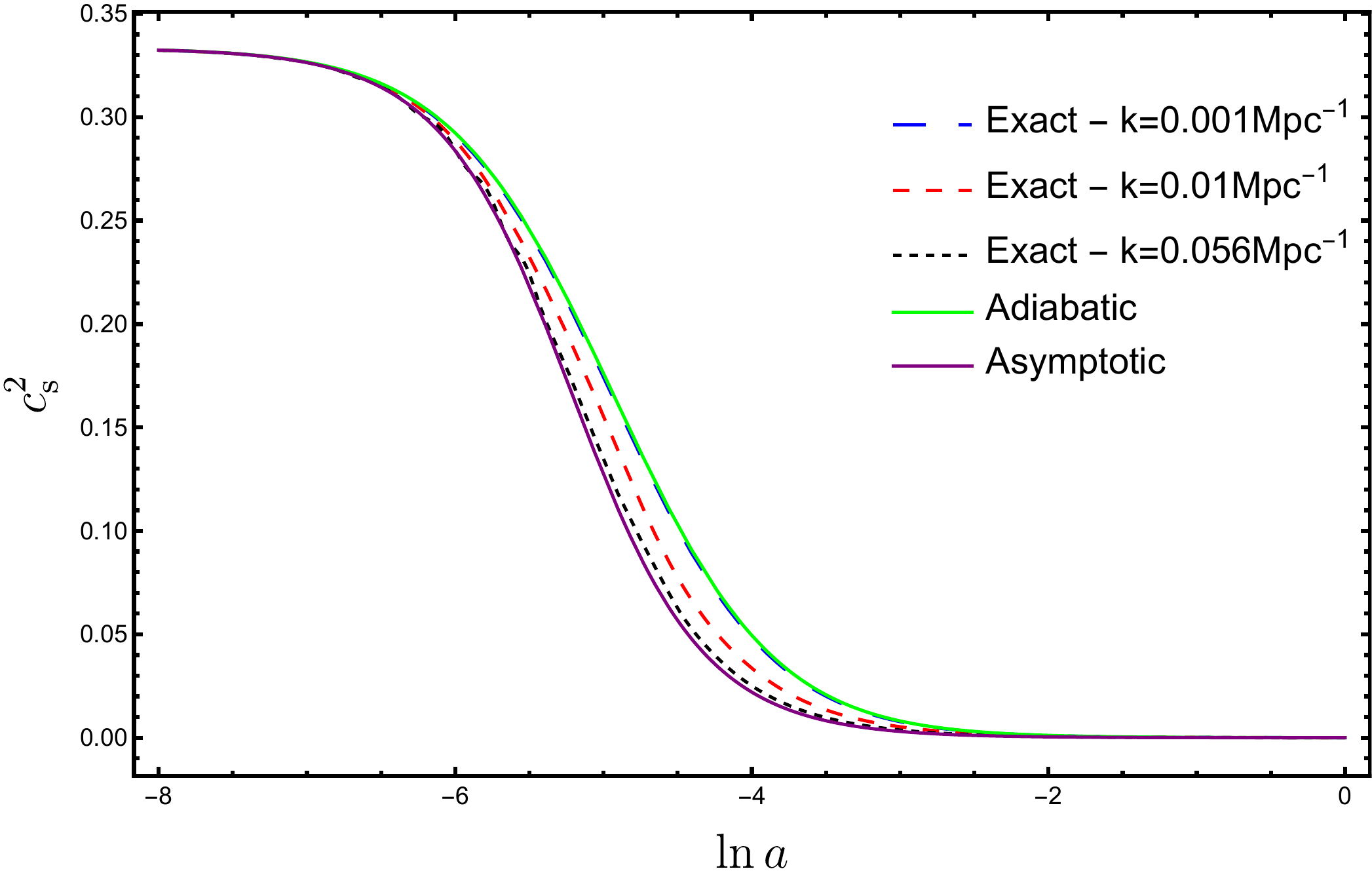}
	\caption{A comparison of both adiabatic and asymptotic sound speeds, as given by Eqs.(\ref{eq:adia}) and (\ref{eq:asp}), with the exact sound speed from the Boltzmann code CLASS in high precision settings (and in the Newtonian gauge) at three different scales and as a function of the scale factor. Here we choose $m_{\nu}=0.1$eV. The green and purple solid curves correspond to the adiabatic and asymptotic sound speeds, respectively. The blue, red and black dashed curves correspond to the exact solutions for $k=10^{-3}$Mpc$^{-1}$, $k=10^{-2}$Mpc$^{-1}$, and $k=5.6 \times 10^{-2}$Mpc$^{-1}$ respectively.}
	\label{fig:sound_speed}
\end{figure}

We are finally ready to write down the approximations we use, after which we  compare to previous approximation schemes and explain the different terms involved. More details can be found in Appendix \ref{sec:app}. The approximations are:
\begin{align}
	& c_{\textrm{s}}^2(a,k) = c_{\textrm{g}}^2(a) + \left[c_{\textrm{asp}}^2(a)-c_{\textrm{g}}^2(a)\right]e^{-\frac{4}{3}\frac{k_{\textrm{fs}}(a)}{k}} \,, \label{eq:approx1} \\ & k^2 \sigma(a,k) = - \frac{2}{5} \frac{k_{\textrm{hor}}(a)}{k}e^{- \frac{k_{\textrm{hor}}(a)}{k}} \frac{c_{\textrm{s}}^2(a,k)}{1+w(a)} k^2 \delta(a,k) + \frac{k}{k_{\textrm{fs}}(a)} e^{-5\frac{k_{\textrm{fs}}(a)}{k}} w^2(a) \theta(a,k) \,. \label{eq:approx2}
\end{align} 
Equation (\ref{eq:approx1}) is a simple interpolation between the two regimes given by Eqs.(\ref{eq:adia}) and (\ref{eq:asp}) \footnote{The precise numerical factors in the exponents of Eqs.(\ref{eq:approx1}) and (\ref{eq:approx2}) are adjusted in such a way as to optimize the fluid approximation.}. It improves on previous approximation schemes in the literature that generally assume an adiabatic sound speed, since we account for deviations from adiabaticity on the small scales (which can be phrased as the presence of an entropy perturbation). This is illustrated by Fig. \ref{fig:sound_speed}, where we compare both the adiabatic and asymptotic expressions [Eqs.(\ref{eq:adia}) and (\ref{eq:asp}) respectively] to the exact sound speed extracted from the Boltzmann code CLASS in high precision settings \footnote{These are the high precision settings we employed in CLASS: \texttt{ncdm\_fluid\_approximation = 3} (this turns off the CLASS fluid approximation), \texttt{Quadrature strategy = 3}, \texttt{Maximum q = 15}, \texttt{Number of momentum bins = 30}, \texttt{l\_max\_ncdm = 30}.} at three different scales and for a neutrino mass $m_{\nu}=0.1$eV. As we move from larger to smaller scales the exact sound speed shifts from the adiabatic to the asymptotic formulas. Note that the adiabatic sound speed overestimates the exact sound speed (a result that holds true in both Newtonian and synchronous gauges).

Next we move on to the anisotropic stress. As we argue in Appendix \ref{sec:app}, we expect it to mostly give an additional contribution to the sound speed at scales that are around the neutrino horizon $k \sim k_{\textrm{hor}}$. This is accomplished by the first term in the right-hand side of Eq.(\ref{eq:approx2}). However, we generally also expect it to give a viscosity type contribution proportional to the divergence of the velocity at small sub-free streaming scales, that turns out to be important for the numerical stability of the fluid equations. This leads to the second term in the right-hand size of Eq.(\ref{eq:approx2}); the precise time dependence is not important, but we choose the equation of state squared because it produces good results. The standard approach in the literature consists in modeling the evolution equation for the shear stress (see \cite{lesgourgues2011cosmic} for a summary) while we introduce an algebraic relation to directly approximate the shear stress in terms of the neutrino density and velocity fields. 

It is important to point out that the approximation in Eq.(\ref{eq:approx2}) is heuristic in the sense that it is only loosely motivated and we do not expect it to accurately reproduce the anisotropic stress. However, the approximation in Eq.(\ref{eq:approx1}) is much more robust and plays a central role in the fluid equations, while Eq.(\ref{eq:approx2}) at least qualitatively accounts for the subtle effects of shear stress on the scales where they are needed. In other words, the precise modeling of the shear stress is not important as long as one is only interested in the neutrino density and velocity fields in the nonrelativistic regime. As we will see in the next section, the fluid approximation benefits from a dramatic increase in accuracy when the dispersive nature of the neutrino fluid is accounted for. 

The approximations in Eqs.(\ref{eq:approx1}) and (\ref{eq:approx2}) are tailored to the Newtonian gauge. An extension to alternative gauges is presented in Appendix \ref{sec:app2}, where we show explicit formulas in the synchronous gauge.

\section{Numerical Results}
\label{sec:num}

We consider the standard fluid equations for the evolution of cosmological perturbations in massive neutrinos species, i.e., Eqs.(\ref{eq:fluid1}) and (\ref{eq:fluid2}), but now with novel scale dependent approximate expressions for the sound speed and shear stress, Eqs.(\ref{eq:approx1}) and (\ref{eq:approx2}) respectively. We are left with a simple closed system of two ordinary differential equations that we refer to as the modified fluid approximation (Modified FA), and solve numerically. 

We extract the Hubble expansion rate and gravitational potentials directly from CLASS so we can just focus on the neutrino species \footnote{In a Boltzmann solver the neutrino species is coupled to all the other species in the universe via the Einstein equations, and so there are additional evolution equations for the Hubble expansion rate (the Friedmann equation) and the gravitational potentials. }. We consider three distinct values for the individual neutrino mass, $m_{\nu}=0.02$eV, $m_{\nu}=0.1$eV and $m_{\nu}=0.5$eV, and solve the evolution equations for 17 wavenumbers ranging from $k_{\textrm{min}}=10^{-3}$ Mpc$^{-1}$ to $k_{\textrm{max}}=10$ Mpc$^{-1}$, equally spaced logarithmically. We set the standard adiabatic initial conditions at super horizon scales, extracting the initial values of neutrino transfer functions directly from CLASS. We similarly evolve the neutrino transfer functions with the CLASS fluid approximation for comparison \cite{lesgourgues2011cosmic}.

\begin{figure}
	\centering
	\includegraphics[width=1\textwidth]{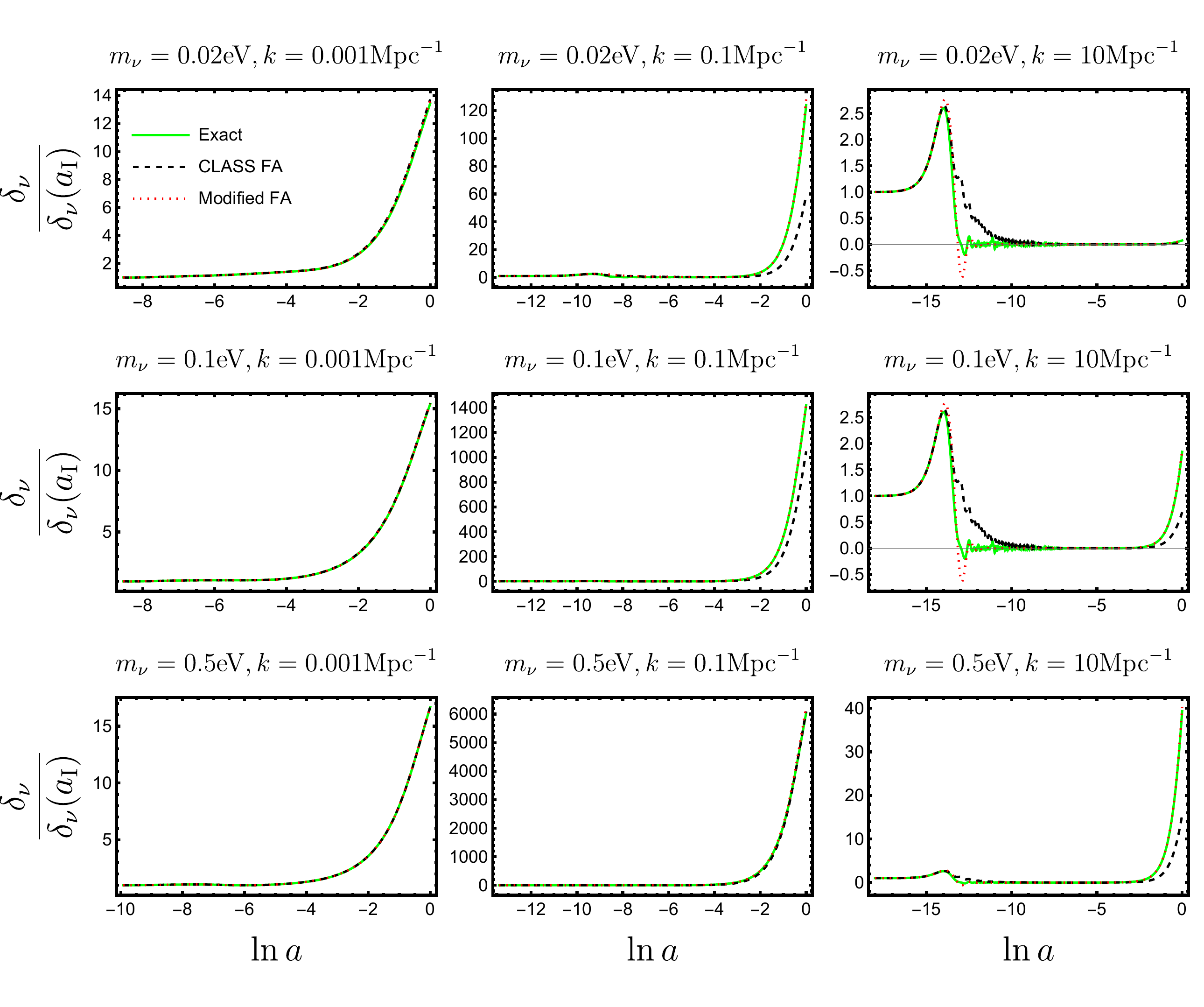}
	\caption{Neutrino density contrast (normalized by its initial value $\delta_{\nu}(a_{\textrm{I}})$, and in the Newtonian gauge) as a function of the scale factor and for varying neutrino mass and scale. The plot compares the modified and CLASS fluid approximations (FA) with the exact solution from CLASS in high precision settings. The solid green curves corresponds to CLASS in high precision settings, black dashed curves to the CLASS FA, and red dotted curves to the Modified FA. }
	\label{fig:time_dependence}
\end{figure}

\begin{figure}
	\centering
	\includegraphics[width=1\textwidth]{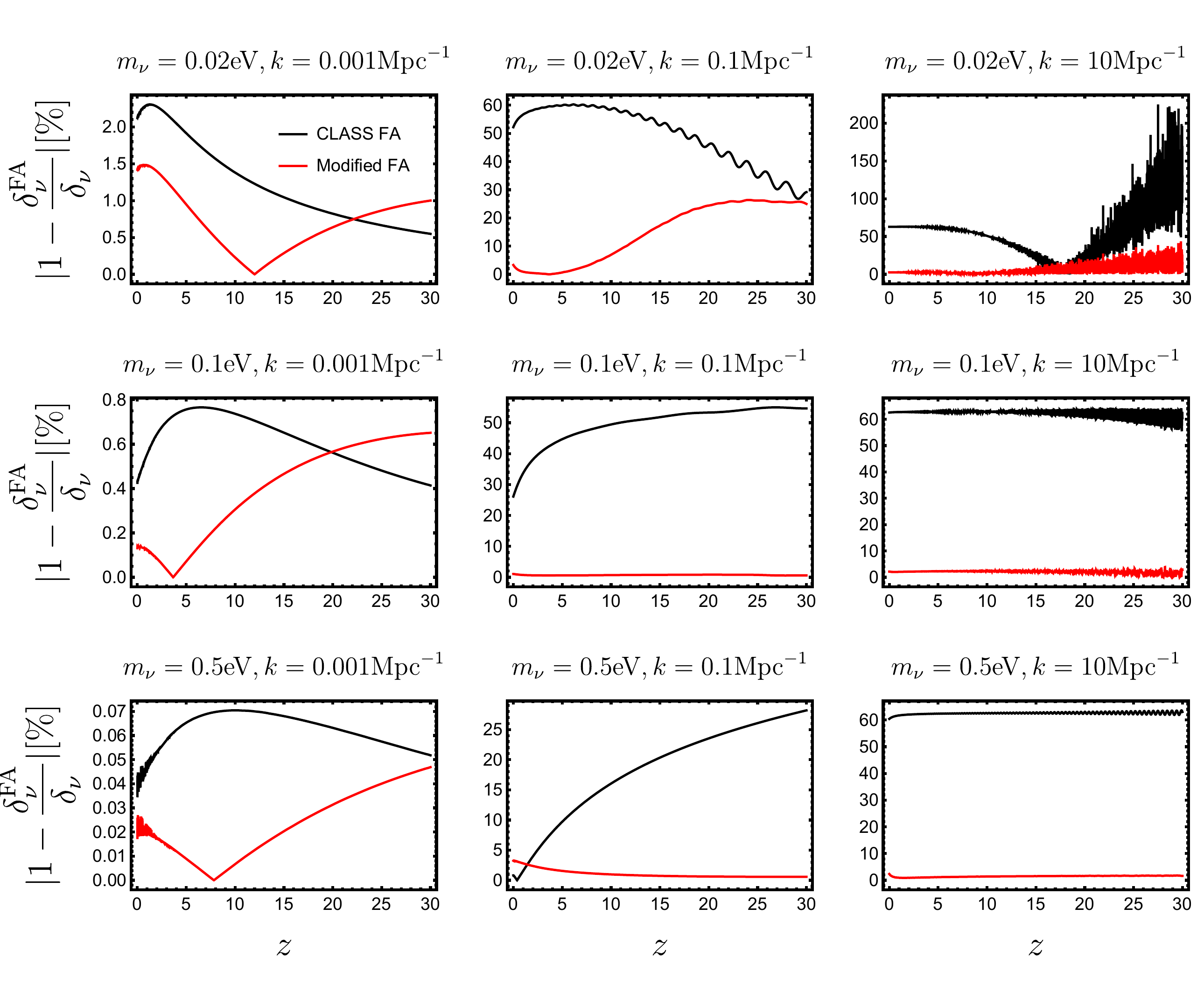}
	\caption{Per cent relative differences in the neutrino density contrast between the exact solution and fluid approximations (in the Newtonian gauge) as a function of redshift (with $z\leq$ 30) for varying neutrino mass and scale. The plot compares the modified and CLASS fluid approximations (FA) with the exact solution from CLASS in high precision settings. The solid black curves corresponds to the CLASS FA and red solid curves to the Modified FA. }
	\label{fig:time_dependence_rel}
\end{figure}

In Fig.\ref{fig:time_dependence} we plot the neutrino density contrast as a function of the scale factor for varying neutrino mass and scale. As one can see from the plot, the Modified FA produces the late-time neutrino growth at intermediate and small scales much more accurately than the CLASS FA. Also, the errors in both fluid approximations can be large in the relativistic regime where the effects of shear stress are significant. In Fig.\ref{fig:time_dependence_rel} we plot the relative differences in the neutrino density contrast, comparing the exact solution to the fluid approximations, as a function of redshift for $z\leq 30$. At large scales the two fluid approximations have a similar performance, but errors are much smaller in the Modified FA when compared to the CLASS FA at intermediate and small scales. Note that, for the Modified FA, the errors are always below a $\lesssim 30\%$ for redshift $z\lesssim 30$ and $\lesssim 5\%$ for redshift $z\lesssim 5$. On the other hand, the CLASS FA can be as much as a factor of two wrong even at $z=0$. We also point out that the fluid approximations are more accurate for larger neutrino masses, which is to be expected since a larger mass implies neutrinos are deeper in the nonrelativistic regime where the contributions from the shear stress can be neglected.

\begin{figure}
	\centering
	\includegraphics[width=1\textwidth]{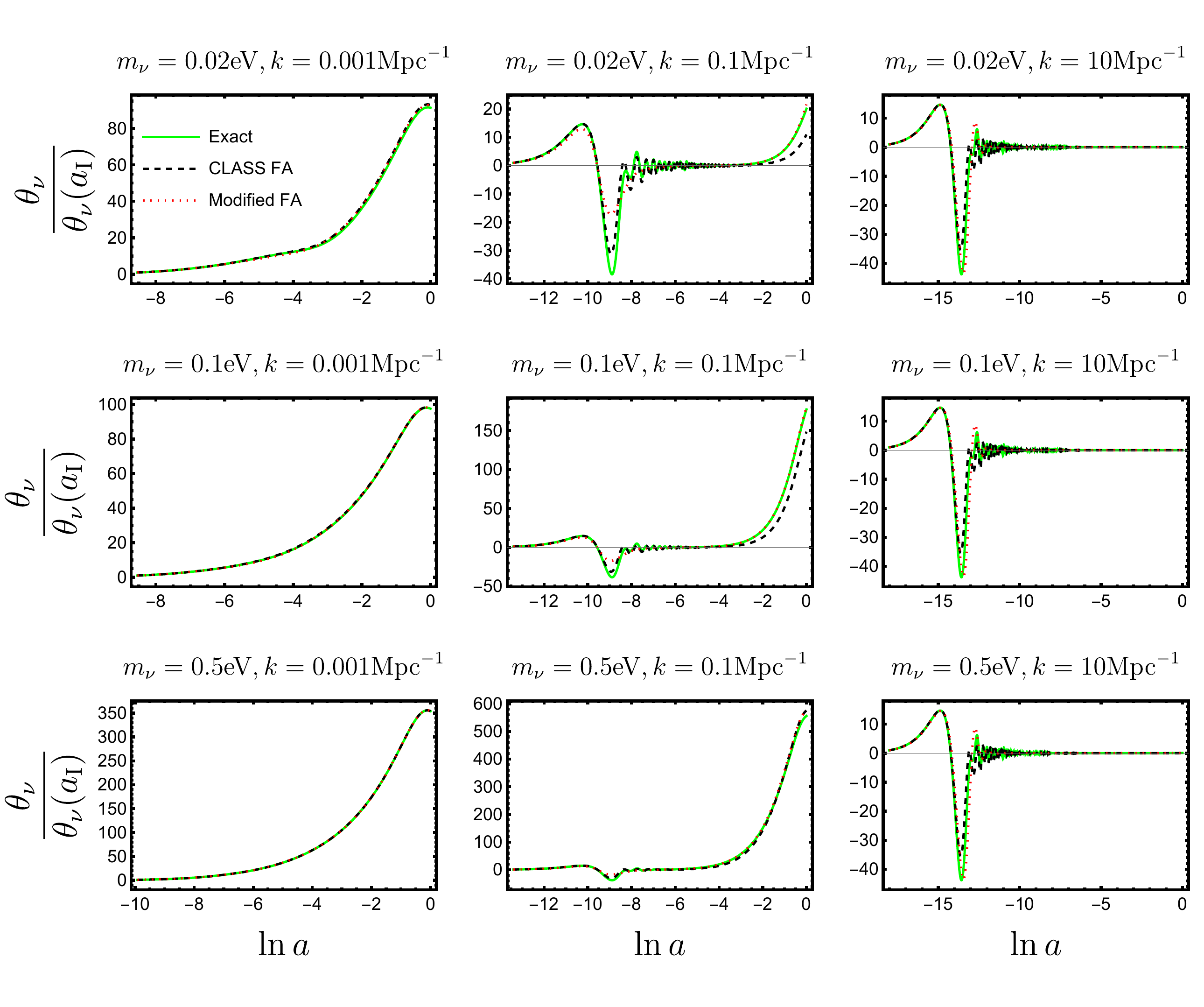}
	\caption{Divergence of the neutrino velocity (normalized by its initial value $\theta_{\nu}(a_{\textrm{I}})$, and in the Newtonian gauge) as a function of the scale factor and for varying neutrino mass and scale. The plot compares the modified and CLASS fluid approximations (FA) with the exact solution from CLASS in high precision settings. The solid green curves corresponds to CLASS in high precision settings, black dashed curves to the CLASS FA, and red dotted curves to the Modified FA. }
	\label{fig:time_dependence_theta}
\end{figure}

\begin{figure}
	\centering
	\includegraphics[width=1\textwidth]{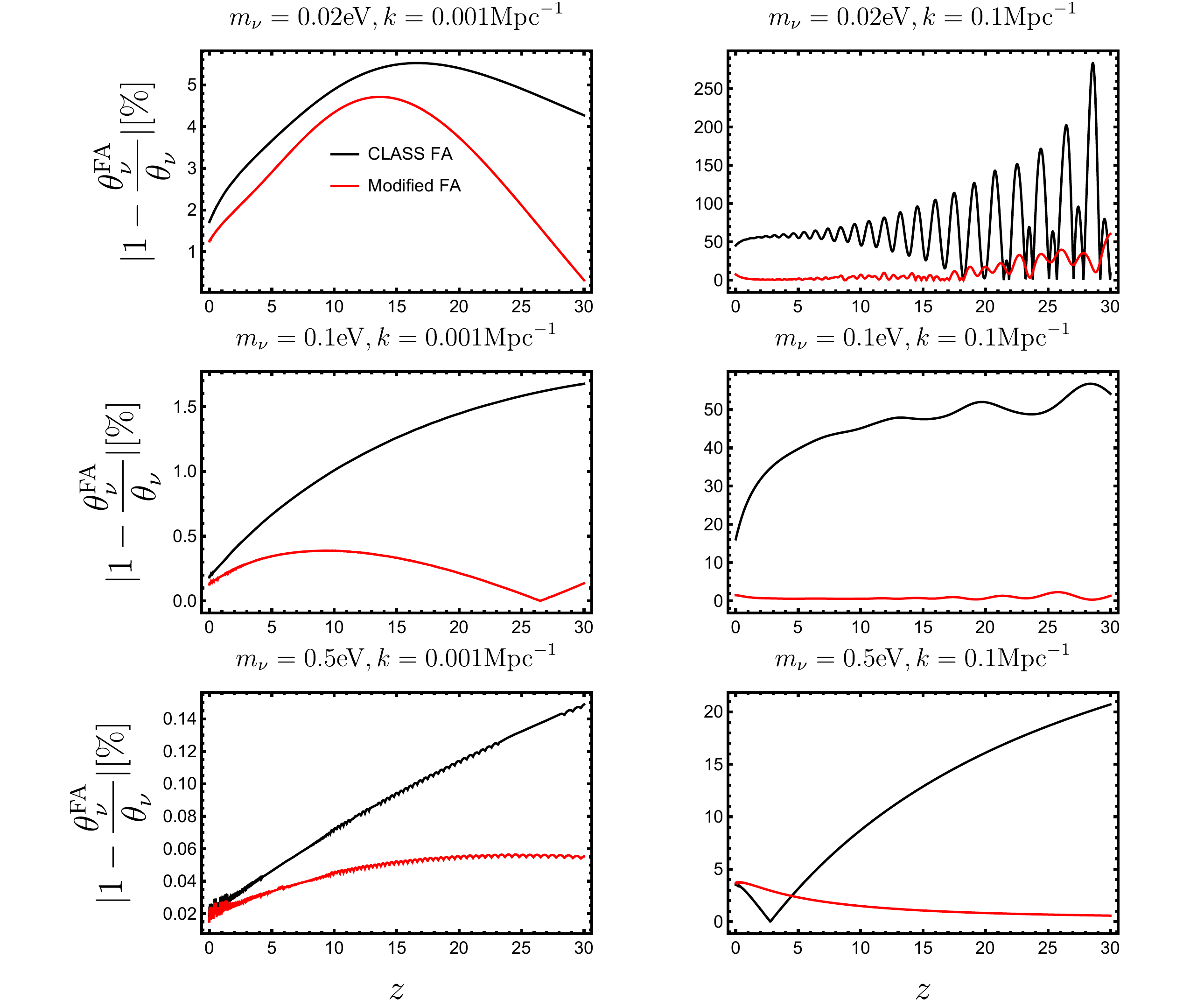}
	\caption{Per cent relative differences in the divergence of the neutrino velocity between the exact solution and fluid approximations (in the Newtonian gauge) as a function of redshift (with $z\leq$ 30) for varying neutrino mass and scale. The plot compares the modified and CLASS fluid approximations (FA) with the exact solution from CLASS in high precision settings. The solid black curves corresponds to the CLASS FA and red solid curves to the Modified FA. }
	\label{fig:time_dependence_rel_theta}
\end{figure}

In Fig.\ref{fig:time_dependence_theta} we plot the divergence of the neutrino velocity as a function of the scale factor for varying neutrino mass and scale, and in Fig.\ref{fig:time_dependence_rel_theta} we plot the relative differences in the divergence of the neutrino velocity, comparing the exact solution to the fluid approximations, as a function of redshift for $z\leq 30$. The divergence of the neutrino velocity displays fast oscillations around zero at the smallest scale and at late times, as can be seen in Fig.\ref{fig:time_dependence_theta}, in which case small phase shifts lead to large relative differences that are insignificant. This is why we choose not to include the $k=10$Mpc$^{-1}$ plots in Fig.\ref{fig:time_dependence_rel_theta}. Once again the Modified FA is significantly more accurate than the CLASS FA at intermediate and small scales, and overall the errors in the divergence of the neutrino velocity are of the same size as the errors in the neutrino density contrast. 

\begin{figure}
	\centering
	\includegraphics[width=1\textwidth]{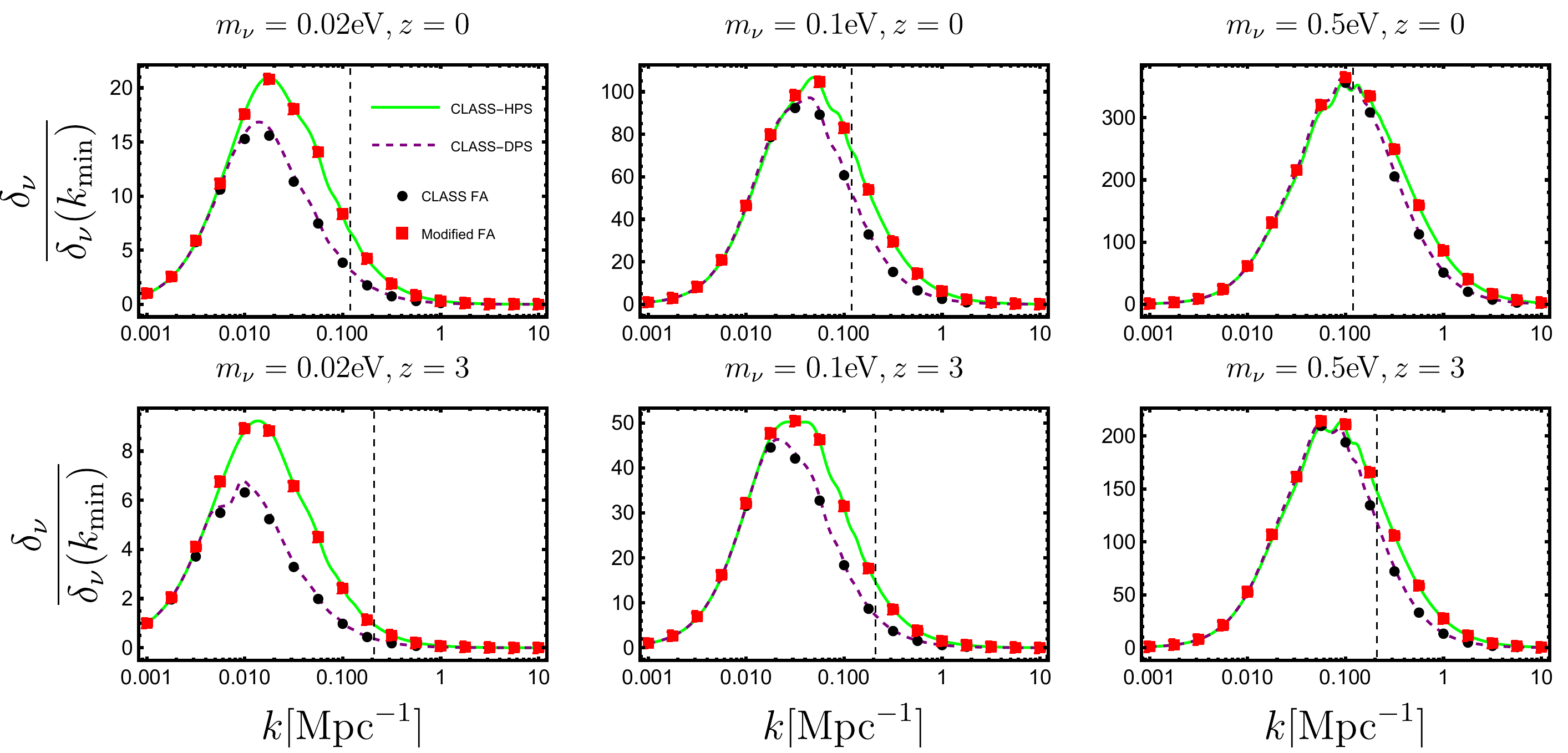}
	\caption{Neutrino density contrast (normalized by its value at the largest scale we consider, $k_{\min}=10^{-3}$Mpc$^{-1}$, and in the Newtonian gauge) as a function of scale and for varying neutrino mass and redshift. The plot compares the modified and CLASS fluid approximations (FA) with the solution from CLASS in both high and default precision settings. The green solid curves corresponds to CLASS in high precision settings, and the dashed purple curves to CLASS in default precision settings. The black round data points are obtained with the CLASS FA and red square data points are obtained with the Modified FA. The black dashed vertical lines correspond to the scale of nonlinearities beyond which the linear perturbation theory is expected to break down.}
	\label{fig:scale_dependence}
\end{figure}

Finally, in Fig.\ref{fig:scale_dependence} we plot the neutrino density contrast as a function of scale for varying neutrino mass and for two values of redshift, $z=0$ and $z=3$. The neutrino density contrast from the Modified FA is in very good agreement with the exact solution from CLASS in high precision settings, with a $\lesssim 5\%$ agreement at redshifts $z \lesssim 5$ at all scales. We expect linear perturbation theory to break down at sufficiently small scales, $k \gtrsim k_{\textrm{NL}}$,  with $k_{\textrm{NL}}$ the scale of nonlinearities, \footnote{We adopt the definition for the scale of nonlinearities based on the root-mean-square linear theory displacement, i.e., $k_{\textrm{NL}}^{-2} = f^2 \int (dk/2\pi^2) P(k)$, with $f=d\log D_{\textrm{L}}/d\log a$ the liner growth rate [in terms of the linear growth factor $D_{\textrm{L}}$(a)] and $P(k)$ the matter power spectrum. This leads to the numerical values of $k_{\textrm{NL}} \approx 0.12$Mpc$^{-1}$ at $z=0$, and $k_{\textrm{NL}} \approx 0.21$Mpc$^{-1}$ at $z=3$, in our reference Planck 2018 cosmology \cite{collaboration2020planck}. An alternative definition is given by the scale where the dimensionless power spectrum becomes unity, which leads to similar numerical values. } which is included as a vertical dashed line in Fig.\ref{fig:scale_dependence}. One can then see from the plot that, and specially for smaller neutrino masses, the Modified FA leads to significant improvements when compared to previous approximations schemes at scales where the linear perturbation theory can be safely applied. 

Since in high precision settings CLASS and CAMB agree to a percent level \cite{dakin2019nuconcept}, we can also conclude that the Modified FA agrees with the Boltzmann solver CAMB as well. Additionally, in Appendix \ref{sec:app2} we argue that the fluid approximation is as accurate in the synchronous gauge as it is in the Newtonian gauge. Furthermore, from Fig.\ref{fig:scale_dependence} the Modified FA is visibly superior to both the CLASS FA and CLASS in default precision settings, specially at scales comparable to, and smaller than, the neutrino free-streaming scale \footnote{In Fig.\ref{fig:scale_dependence}, the fact that the CLASS FA aligns with CLASS in default precision settings is not a coincidence, since in default precision settings CLASS switches from the Boltzmann hierarchy to the CLASS FA after horizon crossing (at $k\tau=15$). }. The Modified FA is then a simple system of two ordinary differential equations that can accurately predict the evolution of linear cosmological neutrino anisotropies at late times.

\section{Conclusion}
\label{sec:conc}

We revisited the fluid approximation for massive neutrinos in linear cosmological perturbation theory, but now accounting for the dispersive nature of the neutrino fluid. In Sec.\ref{sec:fluideqs} we introduced the analytic expressions in Eqs.(\ref{eq:adia}) and (\ref{eq:asp}) for the large and small scales limits of the sound speed respectively (see Fig.\ref{fig:sound_speed}), leading to the novel approximations in Eqs.(\ref{eq:approx1}) and (\ref{eq:approx2}) that can be used to close the fluid Eqs.(\ref{eq:fluid1}) and (\ref{eq:fluid2}).

In Sec. \ref{sec:num} we showed that the resulting modified fluid approximation produces a neutrino transfer function that is in very good agreement with the exact solution from CLASS in high precision settings, achieving a $\lesssim 5\%$ errors for redshifts $z\lesssim 5$, and over scales (at least) in the range $k= (10^{-3}-10)$Mpc$^{-1}$. Furthermore, Figs. \ref{fig:time_dependence_rel}, \ref{fig:time_dependence_rel_theta} and \ref{fig:scale_dependence} show the superiority of the modified fluid approximation when compared to both the  Boltzmann solver CLASS in default precision settings and the CLASS fluid approximation, which corresponds to an order of magnitude improvement over previous approximation schemes that can be as much as a factor of two wrong even at $z=0$.

The modified fluid approximation we propose then offers a simple implementation of massive neutrinos in linear cosmological perturbation theory, being much faster and more versatile than the full Boltzmann hierarchy while delivering accurate neutrino transfer functions at late times. In terms of the sum of neutrino masses $M_{\nu}=\sum_{\nu} m_{\nu}$, the contribution of nonrelativistic neutrinos to the energy density in the Universe is $\Omega_{\nu}h^2  \approx M_{\nu}/93.14$eV \cite{lesgourgues2006massive}. For $M_{\nu} \lesssim 1$eV and $\Omega_{\textrm{m}}h^2 = 0.1424$ \cite{aghanim2021planck}, this leads to a fractional contribution of neutrinos to the total matter of $f_{\nu} \lesssim 7\%$. As a consequence, $\lesssim 5\%$ errors in the neutrino clustering lead to a matter power spectrum that is accurate to the sub-percent level. This estimate is conservative as the effect of neutrino masses in the matter power spectrum comes mostly from the absence of neutrino perturbations at small scales, and not from the clustering of neutrinos itself. We conclude that the Modified FA is sufficient for most applications, with the Boltzmann hierarchy being needed only if one is interested in the neutrino anisotropies at redshift $z>5$, or if the warm dark matter component (neutrinos) has a significant contribution to the total matter. We tested the modified fluid approximation for neutrino masses in the range $m_{\nu}=(0.02-0.5)$eV, but we expect it to remain accurate outside of this range (provided the neutrino mass is large enough for it to become non-relativistic at sufficiently early times and otherwise can be treated as a radiation component). 

In a future work we plan on investigating a reformulation of how massive neutrinos are implemented in the Boltzmann solver CLASS, based on the generalized Boltzmann hierarchy \cite{de2021generalized}, a novel alternative approach to the exact massive neutrino dynamics that can reach sub-percent level accuracy at large and intermediate scales, coupled to our fluid approximation for the small scale dynamics. We will then perform a thorough comparison with standard methods and approximation schemes, and we expect to achieve significant improvements in both accuracy and computation time. Finally, another possible direction of future research is to investigate potential implications of this work to the clustering of neutrinos in the nonlinear regime on the basis of a fluid approach.

\acknowledgements

I would like to thank Zachary Weiner for reading the draft and making great suggestions, and Marilena Loverde for many valuable discussions, feedback on drafts, and overall guidance. I acknowledge support from the Department of Physics and the College of Arts and Sciences at the University of Washington, and the Department of Energy under grant DE-SC0023183.
\appendix 

\section{Analytic calculations and approximations}
\label{sec:app}

We investigate analytically the evolution of cosmological linear perturbations in a massive neutrino species, with the goal of extracting both the large and small scale limits of the sound speed and the qualitative behavior of the anisotropic stress. This study motivates the approximations in Eqs.(\ref{eq:approx1}) and (\ref{eq:approx2}).

We follow \cite{shoji2010massive} and write evolution equations for the phase-space distribution of massive neutrinos. The first step is to split it into a background and perturbation components (working in Fourier space),
\begin{equation}
\label{eq:difunc}
	f(k,q,\mu,\tau) = f_{0}(q)[1+\Psi(k,q,\mu,\tau)] \,,
\end{equation} 
where $k$ the magnitude of the wavevector $\vec{k}$, $q$ the magnitude of the comoving momentum $\vec{q}$, $\mu = \hat{k} \cdot \hat{q}$ is the cosine of the angle between these two vectors, and $\tau$ is the conformal time. Furthermore, the background is a relativistic Fermi-Dirac distribution:
\begin{equation}
	\label{eq:fermidiracapp}
	f_{0}(q) = \frac{g_{\nu}}{e^{\frac{q}{T_{\nu, 0}}}+1} \,,
\end{equation}
with $g_{\nu}=2$ to account for both left-handed neutrinos and right-handed antineutrinos, and $T_{\nu,0} \approx 1.95$K $\approx 1.7 \times 10^{-4}$eV is the neutrino temperature today. We remind the reader that we are working in the Newtonian gauge:
\begin{equation}
	\label{eq:newgaugeapp}
	ds^2 = a^2(\tau)\left[-(1+2\psi)d\tau^2 + (1-2\phi)d\vec{x}^2\right] \,.
\end{equation}
The evolution equation for the perturbation to the phase-space distribution follows from the collisionless Boltzmann equation, and reads \footnote{From this point forward we omit scale, momentum and time dependences when it is convenient to do so. Also, prime denotes a derivative with respect to conformal time.},
\begin{equation}
\label{eq:boltz}
	\Psi' + i\frac{kq}{\epsilon} \mu \Psi = \frac{d\ln f_{0}}{d\ln q} \left(i\frac{k\epsilon}{q} \mu \psi -\phi' \right) \,,
\end{equation}
where $\epsilon(q,\tau) = \sqrt{q^2 +m_{\nu}^2a(\tau)^2}$ is the comoving energy. It is convenient to perform a decomposition of $\Psi(k,q,\mu,\tau)$ into Legendre polynomials, $P_{l}(\mu)$, as follows:
\begin{equation}
\label{eq:multipoles}
	\Psi(k,q,\mu,\tau) = \sum_{l=0}^{\infty} (-i)^l (2l+1) \Psi_{l}(k,q,\tau) P_{l}(\mu) \,,
\end{equation}
where the $\Psi_l(k,q,\tau)$ are called the multipoles of the distribution function. The substitution of Eq.(\ref{eq:multipoles}) into Eq.(\ref{eq:boltz}) yields, after using Legendre polynomial identities,
\begin{align}
	& \Psi'_{0} = -\frac{q}{\epsilon} \Psi_{1} - \frac{d\ln f_{0}}{d\ln q} \phi' \,, \label{eq:dynmulti1}\\ & \Psi'_{1} = \frac{q}{3\epsilon} (\Psi_{0}-2\Psi_{2}) - \frac{\epsilon}{3q} \frac{d\ln f_{0}}{d\ln q} \psi \,, \label{eq:dynmulti2} \\ &  \Psi'_{l} = \frac{q}{(2l+1)\epsilon} [l\Psi_{l-1}-(l+1)\Psi_{l+1}] \  \forall l \geq 2 \,. \label{eq:dynmulti3}
\end{align}
Once the solution to this set of equations is obtained, one can integrate over momenta to get the neutrino stress-energy tensor, and hence the fluid properties involved in the fluid equations (see Eqs.(\ref{eq:fluid1}) and (\ref{eq:fluid2})), as follows \footnote{We remind the reader that we do not employ a $\nu$ subscript when referring to neutrino fluid properties.},
\begin{align}
	& \delta \rho(k,\tau) = a(\tau)^{-4} \int_{0}^{\infty} \frac{dq}{2\pi^2} q^2 \epsilon(q,\tau) f_{0}(q)\Psi_{0}(k,q,\tau) \ , \label{eq:stress1} \\ & \delta P(k,\tau) = \frac{1}{3} a(\tau)^{-4} \int_{0}^{\infty} \frac{dq}{2\pi^2} q^2  \epsilon(q,\tau) \left[\frac{q}{\epsilon(q,\tau)}\right]^2 f_{0}(q)\Psi_{0}(k,q,\tau) \ , \label{eq:stress2} \\ & \rho(1+w)  \theta(k,\tau) = ka(\tau)^{-4} \int_{0}^{\infty} \frac{dq}{2\pi^2} q^3 f_{0}(q)  \Psi_{1}(k,q,\tau) \ \label{eq:stress3} , \\ & \rho(1+w)  \sigma(k,\tau) = \frac{2}{3} a(\tau)^{-4} \int_{0}^{\infty} \frac{dq}{2\pi^2} q^2 \epsilon(q,\tau) \left[\frac{q}{\epsilon(q,\tau)}\right]^2 f_{0}(q)\Psi_{2}(k,q,\tau) \ . \label{eq:stres4}
\end{align}

Here $\rho=\rho(a)$ is the neutrino background energy density as given by Eq.(\ref{eq:fluidprop1}), and $w=w(a)$ is the equation of state parameter from Eq.(\ref{eq:eos}). Furthermore, $\delta \rho(k,\tau)$ is the perturbation to the neutrino energy density, $\delta P(k,\tau)$ the perturbation to the pressure, $\theta(k,\tau)$ is the divergence of the velocity field, and $\sigma(k,\tau)$ is the anisotropic (shear) stress. This sets the stage for an analytic investigation of the sound speed and anisotropic stress.  
\subsection{The sound speed}

The sound speed is defined by:
\begin{equation}
	\label{eq:effspapp}
	c_{\textrm{s}}^2(k,\tau) = \frac{\delta P(k,\tau)}{\delta \rho (k,\tau)} \,,
\end{equation}
and it is both time and scale dependent. From Eqs.(\ref{eq:stress1}) and (\ref{eq:stress2}), one finds that it can be determined from the momentum dependence in the zeroth multipole $\Psi_{0}$, so this is what we investigate next.

From the structure of Eq.(\ref{eq:boltz}) it is convenient to introduce a momentum dependent time variable \footnote{Not to be confused with redshift. Everywhere in the Appendix \ref{sec:app}, the variable $z$ will stand for the distance traveled by neutrino particles over the expansion history as given by Eq.(\ref{eq:ztime}). },
\begin{equation}
\label{eq:ztime}
	dz = \frac{q}{\epsilon} d\tau \implies z(\tau,q) = \int_{0}^{\tau} \frac{q}{\epsilon(q,\tau')} d\tau' \,.
\end{equation}
Since $v \equiv q/\epsilon$ is the peculiar velocity, this is just the distance traveled by the neutrino particles over the entire expansion history. When evaluated at around the peak of the Fermi-Dirac distribution, $q=3T_{\nu,0}$, we denote it by $\bar{z}= z(q=3T_{\nu,0})$. This is analogous to the neutrino horizon scale introduced in Eq.(\ref{eq:hor}). We further define $y=kz$, in terms of which the Eq.(\ref{eq:boltz}) reads,
\begin{equation}
\label{eq:boltznew}
	\frac{\partial \Psi}{\partial y} + i\mu \Psi = \frac{d\ln f_{0}}{d\ln q}\left( ik\frac{\epsilon}{q} \mu \psi - \phi' \right) \,,
\end{equation}
is a first-order ordinary differential equation, whose solution is straightforward to write down:
\begin{equation}
\label{eq:boltzsol}
	\Psi = \frac{d\ln f_{0}}{d\ln q} \int_{0}^{y} dy' \frac{\hat{\epsilon}}{kq} \left(ik \frac{\hat{\epsilon}}{q} \mu \hat{\psi} -\hat{\phi}'\right) e^{-i\mu (y-y')} \,,
\end{equation} 
where we use the hat notation (exemplified by $\hat{\phi}$) to denote time-dependent quantities evaluated at the intermediate time $y'=kz'$ begin integrated.  Additionally, we drop the initial condition terms and send the initial time to zero, i.e., $y_{i} \to 0$, for simplicity. In the regime $y \gg 1$ this is inconsequential as the solution is dominated by the source term and becomes insensitive to the initial conditions. However, this is no longer true in the regime $y \ll 1 $ which corresponds to the largest scales. This does not pose a significant problem for us since we expect the adiabatic sound speed to be recovered on the largest scales (and we will see why that is). A much more important issue is to accurately obtain the small scale limit $y \gg 1$ of the sound speed for which we can safely drop the initial values of the multipoles.

From the multipole expansion in Eq.(\ref{eq:multipoles}) and the orthogonality relations of Legendre polynomlias, we can write:
\begin{equation}
\label{multipolesol}
	\begin{split}
		& \Psi_{l} = \frac{1}{2(-i)^l} \int_{-1}^{1} d\mu \ \Psi P_{l}(\mu) \,.
	\end{split}
\end{equation}
This can be evaluated using Eq.(\ref{eq:boltzsol}), the identity
\begin{equation}
\label{eq:sphericalid}
	e^{-i\mu(y-y')} = \sum_{l=0}^{\infty} (-i)^{l} (2l+1)j_{l}(y-y')P_{l}(\mu) \,,
\end{equation}
where $j_{l}(y)$ are the spherical Bessel functions of the first kind, and the orthogonality relations of Legendre polynomials, to yield:
\begin{equation}
\label{eq:solagain}
	\Psi_{l} = \frac{d\ln f_{0}}{d\ln q} \int_{0}^{y} dy' \left\{ \left(\frac{\hat{\epsilon}}{q}\right)^2 \hat{\psi} \left[\frac{l+1}{2l+1} j_{l+1}(y-y')-\frac{l}{2l+1} j_{l-1}(y-y') \right] - \frac{\hat{\epsilon}}{kq} \hat{\phi}' j_{l}(y-y')\right\}  \,.
\end{equation}
We now use the following recurrence relation of the spherical Bessel functions,
\begin{equation}
\label{eq:rrbessel}
	lf_{l-1}(y) - (l+1)f_{l+1}(y) = (2l+1)f_{l}'(y) \,, 
\end{equation}
and integrate the first term in the right-hand side of Eq.(\ref{eq:solagain}) by parts to arrive at: 
\begin{equation}
\label{eq:solagain2}
	\Psi_{l} = \frac{d\ln f_{0}}{d\ln q} \left\{ j_{l}(0) \left(\frac{\epsilon}{q}\right)^2 \psi - j_{l}(y) \psi_{i} - \int_{0}^{y} dy' \frac{\hat{\epsilon}}{kq} \left[ \left(\frac{\hat{\epsilon}}{q}\right)^2 \hat{\psi} +\hat{\phi} \right]'  j_{l}(y-y')\right\} \,.
\end{equation}

Now the crucial observation is that this greatly simplifies in the limit $y \gg 1$ for $l=0$, since then $j_{0}(y-y')$ peaks at $\Delta y=y-y'=0$, and goes to zero in the limit $\Delta y=k(z-z')=k\Delta z = y \Delta z/z \gg 1$, and hence $\Delta z/z \gg 1/y$ is a small number. As a consequence, simply evaluating the integrand at the final time should give a very good approximation and indeed eventually approach the exact solution in the asymptotic regime $x \to \infty$. Indeed, this has been previously investigated and exploited to generate a fluid approximation, in the special case of the massless limit, i.e., for a radiation component \cite{blas2011cosmic}. Using $j_{0}(0)=1$ and $j_{0}(y) \approx 0$ for $y \gg 1$ leads to,
\begin{equation}
\label{eq:aspsol}
	\begin{split}
	\Psi_{0} & \approx \frac{d\ln f_{0}}{d\ln q} \left\{ \left(\frac{\epsilon}{q}\right)^2 \psi - \frac{\epsilon}{kq} \left[ \left(\frac{\epsilon}{q}\right)^2 \psi +\phi \right]' \int_{0}^{y} dy'   j_{0}(y-y') \right\} \\ & \approx \frac{d\ln f_{0}}{d\ln q} \left\{ \left(\frac{\epsilon}{q}\right)^2 \psi + \frac{\pi}{2} \frac{\epsilon}{kq} \left[ \left(\frac{\epsilon}{q}\right)^2 \psi +\phi \right]' \right\} \\ & \approx \frac{d\ln f_{0}}{d\ln q} \left(\frac{\epsilon}{q}\right)^2 \psi \,, 
	\end{split}
\end{equation}
where we used, 
\begin{equation}
\label{eq:asptrick}
	 \int_{0}^{y} dy'   j_{0}(y-y') \approx -\int_{0}^{\infty} dy j_{0}(y) = -\frac{\pi}{2} \,,
\end{equation}
and from the second to the third line in Eq.(\ref{eq:aspsol}) we dropped a term that scales like $1/k$ and hence becomes negligible in the limit $y \to \infty$. Indeed, that term is the next to leading order correction to the asymptotic formula. 

The final formula in the third line of Eq.(\ref{eq:aspsol}) has a straightforward interpretation. Neutrino fluctuations with (comoving) wavenumber $k$ have a characteristic time scale, $t \sim a/k\sigma_{\nu} \sim (k_{\textrm{fs}}/k)H^{-1}$, which is much smaller than a Hubble time $ \sim 1/H$, at sub-free streaming scales ($k \gg k_{\textrm{fs}}$). We can then consider a static limit of the Boltzmann equation where the expansion of the universe can be taken as slow. In fact, one can check that simply dropping the derivative terms in Eq.(\ref{eq:boltz}) immediately leads to our final formula in the third line of Eq.(\ref{eq:aspsol}).  

It is now straightforward to obtain the asymptotic formula for the sound speed, using Eqs.(\ref{eq:stress1}),(\ref{eq:stress2}) and (\ref{eq:effspapp}). We first explicitly evaluate the perturbation to the energy density, using Eq.(\ref{eq:aspsol}),
\begin{equation}
\label{eq:enedens}
	\begin{split}
		 \delta \rho & \approx a^{-4} \psi \int_{0}^{\infty} \frac{dq}{2\pi^2} q^3 \epsilon(q,a) \left[\frac{\epsilon(q,a)}{q}\right]^2 \frac{df_{0}(q)}{dq} \\ & = -3\rho(a)[1+\lambda(a)] \psi \implies \delta \approx -3[1+\lambda(a)] \psi \,,
	\end{split}
\end{equation}
where from the first to the second line we integrate by parts and introduce the quantity $\lambda(a)$ as:
\begin{equation}
	\label{eq:lambdaapp}
	\rho(a)\lambda(a) = \frac{1}{3} a^{-4} \int_{0}^{\infty} \frac{dq}{2\pi^2} q^2 \epsilon(q,a) \left[\frac{\epsilon(q,a)}{q}\right]^2 f_{0}(q) \,.
\end{equation}
We can obtain the asymptotic expression for the neutrino pressure in a way that is completely analogous to the density and leads to a sound speed of, using the Eq.(\ref{eq:effspapp}),
\begin{equation}
	\label{eq:aspapp}
	c_{\textrm{asp}}^2(a) = \frac{1}{3} \frac{1+w(a)}{1+\lambda(a)} \,.
\end{equation}

Before moving on to the opposite large scale regime of the sound speed, it is instructive to investigate the non-relativistic limit of Eq.(\ref{eq:enedens}) where $\epsilon \approx ma$ and $\lambda \gg 1$, such that,
\begin{equation}
\label{eq:nrlimit}
	\delta \approx -3\lambda(a)\psi \approx - \frac{1}{\rho} \frac{\ln 4}{2\pi^2} \frac{m_{\nu}^3 T_{\nu,0}}{a} \psi \,,
\end{equation}
after using the Eq.(\ref{eq:fermidiracapp}). In a universe with matter background energy density $\rho_{\textrm{m}}(a)$ and matter density contrast $\delta_{\textrm{m}}(a)$, the Poisson equation reads:
\begin{equation}
\label{eq:poisson}
	k^2 \psi = -4\pi G a^2 \rho_{\textrm{m}} \delta_{\textrm{m}} \,,
\end{equation}
where $G$ is Newton's gravitational constant. Using this, the Friedmann equation,
\begin{equation}
\label{eq:friedmann}
	\mathcal{H}^2 = \frac{8\pi G}{3} a^2 \frac{\rho_{\textrm{m}}}{\Omega_{\textrm{m}}} \,,
\end{equation}
with $\Omega_{\textrm{m}}(a)$ the fractional contribution of matter to the total energy density and also evaluating explicitly the Eq.(\ref{eq:fluidprop1}) for $\rho(a)$ in the non-relativistic regime, we can write the Eq.(\ref{eq:nrlimit}) as:
\begin{equation}
	\label{eq:solfinal}
		\delta = \left(\frac{k_{\textrm{fs}}}{k}\right)^2 \delta_{\textrm{m}}\,, 
\end{equation}
where, 
\begin{equation}
	\label{eq:fsapp}
	k_{\textrm{fs}}(a) = \sqrt{\frac{3}{2} \Omega_{\textrm{m}}(a)} \frac{\mathcal{H}(a)}{\sigma_{\nu}(a)} \,, 
\end{equation}
is (proportional to) the distance that neutrinos travel over the course of one expansion time $t \sim 1/\mathcal{H}$, i.e., $\lambda_{\textrm{fs}} \sim 1/k_{\textrm{fs}} \sim \sigma_{\nu}t \sim \sigma_{\nu}/\mathcal{H}$, with,
\begin{equation}
	\label{eq:vdsapp}
	\sigma_{\nu}(a) = \sqrt{\frac{3\zeta(3)}{\ln 4}} \frac{T_{\nu,0}}{m_{\nu}a} \,,
\end{equation}
the expression for the neutrino velocity dispersion, chosen in such a way as to produce the Eq.(\ref{eq:solfinal}) with no additional coefficients. Indeed, the small scale result in Eq.(\ref{eq:solfinal}) is well-known in the literature \cite{ringwald2004gravitational} and the next to leading order correction in the second line of Eq.(\ref{eq:aspsol}) leads to a contribution that scales as $\sim (k_{\textrm{fs}}/k)^3$.

We are now ready to obtain an expression for the sound speed in the large scale regime $k \ll k_{\textrm{fs}}$. For this we go back to the Eq.(\ref{eq:solagain2}) in the case of $l=0$, but now assume $\Delta y \leq y \ll 1$ such that $j_{0}(\Delta y) \approx 1$ and we obtain,
\begin{equation}
\label{eq:largescale}
	\int_{0}^{y} dy' \frac{\hat{\epsilon}}{kq} \left[ \left(\frac{\hat{\epsilon}}{q}\right)^2 \hat{\psi} +\hat{\phi} \right]'  j_{0}(y-y') \approx \left(\frac{\epsilon}{q}\right)^2 \psi +\phi -\psi_{i} - \phi_{i} \,,
\end{equation}
upon using $d\tau' = dy' (\hat{\epsilon}/kq)$ and assuming the neutrinos to be relativistic at the initial time. This yields:
\begin{equation}
\label{eq:largescale2}
	\Psi_{0} \approx (\phi_{i}-\phi) \frac{d\ln f_{0}}{d\ln q} \,. 
\end{equation}

As we argued previously, we do not necessarily expect the Eq.(\ref{eq:largescale2}) to give an accurate approximation to the zeroth multipole on the large scales, since we dropped the initial values of the multipoles that play a role in this regime. However, the statement that the momentum dependence of $\Psi_{0}$ is set by the log derivative of the background distribution function (the so-called separable ansatz), still holds true at sufficiently large scales as it follows from separate universe arguments (see \cite{brinckmann2022confronting, oldengott2017interacting}). We may now proceed to obtain an expression for the sound speed as before, using Eqs.(\ref{eq:stress1}), (\ref{eq:stress2}), (\ref{eq:effspapp}) and (\ref{eq:largescale2}). We get the so-called adiabatic sound speed,
\begin{equation}
	\label{eq:adiaapp}
	c_{\textrm{g}}^2(a) = \frac{P'(a)}{ \rho'(a)} \,,
\end{equation}
after using the Eqs.(\ref{eq:fluidprop1}) and (\ref{eq:fluidprop2}) for the neutrino energy density $\rho(a)$, and pressure $P(a)$, respectively. Also, the derivatives with respect to conformal time can be obtained via the relation $\epsilon' = \mathcal{H} \epsilon[1-(q/\epsilon)^2]$.

Now equipped with both $k \gg k_{\textrm{fs}}$ and $k \ll k_{\textrm{fs}}$ limits of the sound speed in the Newtonian gauge, as given by Eqs.(\ref{eq:aspapp}) and (\ref{eq:adiaapp}) respectively, we see that an interpolation such as the one provided by the Eq.(\ref{eq:approx1}) should yield a good approximation to the scale-dependent sound speed. 

\subsection{Anisotropic stress}

After obtaining an expression for the sound speed in both the $k/k_{\textrm{fs}} \gg 1$ and $k/k_{\textrm{fs}} \ll 1$ regimes, we would like to develop a qualitative understanding on the behavior of the anisotropic stress following the discussion in \cite{shoji2010massive}. 

Combining the Eqs.(\ref{eq:dynmulti1}) and (\ref{eq:dynmulti3}) for $l=2$, we obtain,
\begin{equation}
\label{eq:shear1}
	\left(\Psi_{2} + \frac{2}{5} \Psi_{0} +\frac{2}{5} \phi \frac{d\ln f_{0}}{d\ln q} \right)' = -\frac{3}{5} \frac{q}{\epsilon} \Psi_{3} \,.
\end{equation}
Since our goal is to build intuition about the anisotropic stress we can make some simplifying approximations. The first is to set $\phi'\approx 0$, which is always a good approximation on small scales, and exact in a matter dominated universe. The second will be to drop the initial values of the multipoles, which is the same approximation we applied in the last subsection. This leads to,
\begin{equation}
\label{eq:shear2}
	\Psi_{2} \approx -\frac{2}{5} \Psi_{0} - \frac{3}{5} \int_{0}^{y} dy' \frac{\hat{\Psi}_{3}}{k} \ ,
\end{equation}
where we used the Eq.(\ref{eq:ztime}), $y=kz$, and the hat notation to denote time-dependent quantities evaluated at the intermediate time $y'$ to be integrated. Using the Eqs.(\ref{eq:stress2}) and (\ref{eq:stres4}), we conclude that the first term in the right-hand side of Eq.(\ref{eq:shear2})  generates the following contribution to the anisotropic stress,
\begin{equation}
\label{eq:shear3}
	\sigma \supset -\frac{4}{5} \frac{c_{\textrm{s}}^2}{1+w} \delta \ ,
\end{equation}
which in the fluid equations looks like an additional contribution to the sound speed.  However, we expect that setting $\Psi_{3} =0$, which is equivalent to dropping the second term in the right-hand side of Eq.(\ref{eq:shear2}), is only a reasonable approximation when $y=kz \lesssim 3$ such that the shear acts like a contribution to the sound speed at scales that are around the horizon $k \sim 1/\bar{z} \sim k_{\textrm{hor}}$. In the regime $y \gg 1$ we expect the second term in the right-hand side of Eq.(\ref{eq:shear2}) to dominate, generally giving a viscosity-type contribution to the shear stress at scales that are comparable to, or smaller than, the free-streaming scale ($k \gtrsim k_{\textrm{fs}}$). These considerations motivate the heuristic approximate expression in Eq.(\ref{eq:approx2}).

\section{Alternative gauges}
\label{sec:app2}

In the main text we work solely in the conformal Newtonian gauge for simplicity. In this section we extend our fluid approximation to a general alternative gauge under the example of the synchronous gauge due to its usage in Boltzmann solvers, for concreteness. We follow \cite{ma1995cosmological} where more details can be found.

Small scalar perturbations to the FLRW universe are given by, in the synchronous gauge: 
\begin{equation}
\label{eq:sync}
	ds^2 = a^2(\tau)[-d\tau^2 + (\delta_{ij}+h_{ij})dx^i dx^j] \,,
\end{equation}
where $\delta_{ij}$ is the Kronecker symbol, and $h_{ij}( \tau, \vec{x})$ can be decomposed (in Fourier space), in terms of two metric perturbations $h(\tau, \vec{k})$ and $\eta(\tau, \vec{k})$, as follows:
\begin{equation}
\label{eq:syncpot}
	h_{ij}(\tau, \vec{x}) = \int \frac{d^3 \vec{k}}{(2\pi)^3} e^{i\vec{k} \cdot \vec{x}} \left[ \hat{k}_{i} \hat{k}_{j} h(\tau,\vec{k}) + (\hat{k}_{i} \hat{k}_{j} - \frac{1}{3} \delta_{ij})6\eta(\tau,\vec{k}) \right] \,,
\end{equation}
with $\hat{k}_{i} = \vec{k}_{i}/k_{i}$ normalized to unit length. In other gauges the metric will similarly be written in terms of two other scalar metric perturbations (such as $\psi$ and $\phi$ in the Newtonian gauge). 

In the synchronous gauge the fluid equations that follow from stress-energy conservation read,
\begin{align}
	& \tilde{\delta}' = -(1+w)\left(\tilde{\theta}+\frac{h'}{2}\right) -3\mathcal{H}(\tilde{c}_{s}^2-w)\tilde{\delta} \,,  \label{eq:fluidsync1}\\ & \tilde{\theta}' = - \mathcal{H}(1-3w) \tilde{\theta} -\frac{w'}{1+w} \tilde{\theta} + \frac{\tilde{c}_{s}^2}{1+w}k^2 \tilde{\delta} -k^2 \tilde{\sigma}  \,. \label{eq:fluidsync2}
\end{align}
Here $\tilde{\delta}$ is the neutrino density contrast, $\tilde{\theta}$ is the divergence of the velocity field, $\tilde{\sigma}$ is the anisotropic stress, and the pressure perturbation term $\delta \tilde{P}$ is parameterized by the sound speed $\tilde{c}_{s}^2$ as in the main text,
\begin{equation}
\label{eq:soundspeed}
	\frac{\delta \tilde{P}}{\rho} = \frac{\delta \tilde{P}}{\delta \tilde{\rho}} \tilde{\delta} = \tilde{c}_{s}^2 \tilde{\delta} \,,
\end{equation}
where from now on tilded quantities are in the alternative gauge (which in our example is the synchronous gauge) and untilded quantities are in the Newtonian gauge. These are the neutrino fluid properties that are involved in the fluid equations. Also $\rho(a)$ and $P(a)$ are the background density and pressure respectively, and $w=P/\rho$ is the equation of state. 

Starting from Newtonian gauge coordinates $x^{\mu}$, we can apply a gauge transformation $x^{\mu}  \to  \tilde{x}^{\mu}= x^{\mu}  +d^{\mu}(x^{\nu})$ in order to arrive at an arbitrary gauge. For scalar perturbations, the $d^{\mu}$ can be decomposed into time and spatial components as,
\begin{align}
	& d^{0}(x^{0},\vec{x}) = \alpha(\tau, \vec{x}) \ , \label{eq:gauge1}\\ &  \vec{d}(x^{0},\vec{x}) = \vec{\nabla} \beta(\tau, \vec{x})   \ . \label{eq:gauge2}
\end{align}
In the example of the synchronous gauge, these are given by (in Fourier space),
 \begin{align}
 	& \beta = \frac{1}{2k^2} (h+6\eta) \,, \label{eq:gaugesync1}\\ &  \alpha = \beta'   \,. \label{eq:gaugesync2}
 \end{align}
In general, $\alpha$ and $\beta$ will be given in terms of the two metric perturbations in the alternative gauge. The transformation laws for the fluid properties in the Eqs.(\ref{eq:fluidsync1}) and (\ref{eq:fluidsync2}) (in any gauge, here illustrated in the synchronous gauge) follow from the covariant transformation law satisfied by the energy-momentum tensor, and read
\begin{align}
	& \tilde{\delta} = \delta - \alpha \frac{\rho'}{\rho} \,, \label{eq:gaugetrans1}\\ &  \tilde{\theta} = \theta - \beta' k^2   \ , \label{eq:gaugetrans2} \\ & \delta \tilde{P} = \delta P - \alpha P' \ , \\ & \tilde{\sigma} = \sigma \,.
\end{align}
Our approximations in Eqs.(\ref{eq:approx1}) and (\ref{eq:approx2}), suitable to the Newtonian gauge, can then be straightforwardly mapped into an arbitrary alternative gauge as follows:
\begin{align}
	& \frac{\delta \tilde{P}}{\rho} = c_{\textrm{s}}^2 \tilde{\delta} + \alpha (c_{\textrm{s}}^2-c_{\textrm{g}}^2) \frac{\rho'}{\rho} \,, \label{eq:approxsync1}\\ &  k^2 \tilde{\sigma} = - \frac{2}{5} \frac{k_{\textrm{hor}}}{k}e^{- \frac{k_{\textrm{hor}}}{k}} \frac{c_{\textrm{s}}^2}{1+w} k^2 \left(\tilde{\delta} + \alpha \frac{\rho'}{\rho}\right) + \frac{k}{k_{\textrm{fs}}} e^{-5\frac{k_{\textrm{fs}}}{k}} w^2 \left(\tilde{\theta} + \beta' k^2 \right)  \,, \label{eq:approxsync2}
\end{align}
where, 
\begin{equation}
\label{eq:csagain}
c_{\textrm{s}}^2 = c_{\textrm{g}}^2 + (c_{\textrm{asp}}^2-c_{\textrm{g}}^2)e^{-\frac{4}{3}\frac{k_{\textrm{fs}}}{k}} \,,
\end{equation}
is the approximate expression for the scale-dependent sound speed in the Newtonian gauge. Also, the adiabatic $c_{\textrm{g}}^2$ and asymptotic $c_{\textrm{asp}}^2$ sound speeds are given by Eqs.(\ref{eq:adia}) and (\ref{eq:asp}) respectively, and the neutrino free-streaming $k_{\textrm{fs}}$ and horizon $k_{\textrm{hor}}$ scales are given by Eqs.(\ref{eq:fs}) and (\ref{eq:hor}), respectively.
 
We were not able to directly test the fluid Eqs.(\ref{eq:fluidsync1}) and (\ref{eq:fluidsync2}), with the approximations in Eqs.(\ref{eq:approxsync1}) and (\ref{eq:approxsync2}) in the synchronous gauge, due to the fact that the Boltzmann solver CLASS does not output the synchronous gauge metric perturbations $h$ and $\eta$ as a function of the scale factor for a given scale $k$. However, we were able to extract the function $\alpha$, applying the Eq.(\ref{eq:gaugetrans1}) with both the Newtonian and synchronous gauge exact solutions (obtained from CLASS in high precision settings). This allowed us to use the fluid equations to obtain the approximate solution in the Newtonian gauge and then transform that into the synchronous gauge, which is mathematically equivalent to solving the fluid equations in the synchronous gauge. When comparing to the exact synchronous gauge transfer functions we obtain a level of accuracy which is the same as observed in the Newtonian gauge, hence verifying that the fluid approximation works as well in the synchronous gauge as it does in the Newtonian gauge. 

\bibliography{fluid_approximation.bib}

\end{document}